\documentclass[authoryear,final,5p,times,twocolumn]{elsarticle}

 \usepackage{graphicx}

\usepackage{amssymb}

\journal{New Astronomy}

\def\aap{A\&A}

\def\aj{AJ}
\def\apj{ApJ}

\def\apjl{ApJ}

\def\mnras{MNRAS}

\def\aapr{A\&ARev}
\def\newa{NewA}
\def\astrobj#1{#1}

\begin{document}

\begin{frontmatter}



\title{The XMM-Newton view of the yellow hypergiant IRC\,+10420 and its surroundings\tnoteref{xmm}}
\tnotetext[xmm]{Based on observations with XMM-Newton, an ESA Science Mission with instruments and contributions directly funded by ESA Member states and the USA (NASA).}

\author{M. De Becker}
\ead{debecker@astro.ulg.ac.be}

\author{D. Hutsem\'ekers\fnref{fnrs}}
\ead{hutsemekers@astro.ulg.ac.be}
\fntext[fnrs]{Senior Research Associate FRS/FNRS (Belgium)}

\author{E. Gosset\fnref{fnrs}}
\ead{gosset@astro.ulg.ac.be}

\address{Department of Astrophysics, Geophysics and Oceanography, University of Li\`ege, 17, All\'ee du 6 Ao\^ut, B5c, B-4000 Sart Tilman, Belgium}

\begin{abstract}
Among evolved massive stars likely in transition to the Wolf-Rayet phase, IRC\,+10420 is probably one of the most enigmatic. It belongs to the category of yellow hypergiants and it is characterized by quite high mass loss episodes. Even though IRC\,+10420 benefited of many observations in several wavelength domains, it has never been a target for an X-ray observatory. We report here on the very first dedicated observation of IRC\,+10420 in X-rays, using the XMM-Newton satellite. Even though the target is not detected, we derive X-ray flux upper limits of the order of 1--3\,$\times$\,10$^{-14}$\,erg\,cm$^{-2}$\,s$^{-1}$ (between 0.3 and 10.0\,keV), and we discuss the case of IRC\,+10420 in the framework of emission models likely to be adequate for such an object. Using the Optical/UV Monitor on board XMM-Newton, we present the very first upper limits of the flux density of IRC\,+10420 in the UV domain (between 1800 and 2250\,\AA\, and between 2050 and 2450\,\AA\,). Finally, we also report on the detection in this field of 10 X-ray and 7 UV point sources, and we briefly discuss their properties and potential counterparts at longer wavelengths. 
\end{abstract}


\end{frontmatter}

\section{Introduction} \label{intro}

The so-called standard evolution scheme of massive OB-stars considers that they evolve to the Wolf-Rayet (WR) phase. This transition is still poorly understood, mainly because such intermediate phases are characterized by short time-scales compared to main-sequence time-scales. In addition to the scarcity of massive stars (considering the steepness of the stellar mass function), the brevity of this transition makes such objects very rare. For instance, evolved O-type stars such as Of$^+$ supergiants present some spectral similarities with WR stars \citep[e.g.][]{contitrans}, such as very strong and broad emission lines in the infrared. One can also consider the case of Luminous Blue Variables (LBV): very bright blue stars that undergo sudden eruptions with high mass loss rates \citep[e.g.][]{davidsonlbv,notalamers}. The nature of such objects (P\,Cygni, AG\,Car...) is not yet elucidated, but they are most probably evolved massive stars in transition to the WR stage. On the other hand, even more rare objects have been found in our Galaxy, also with very large luminosities but not in the blue part of the Hertzsprung-Russell (H-R) diagram: the so-called yellow hypergiants. In their evolution, these very massive stars enter a temperature range (6000--9000\,K) of increased dynamical instability, where high-mass loss episodes occur \citep{humphreys2002}. Until a couple of years ago, only two yellow hypergiants were reputed to be accompanied by circumstellar shells, and believed to be located very close to the beginning of the Wolf-Rayet phase: \astrobj{IRC+10420} and \astrobj{HD179821} \citep{oudmaijer2009}. More recently \citet{lagadec2011} suggested that \astrobj{IRAS 17163-3907} (Hen 3-1379) might be the third member of this scarce category, and \citet{clark2013} proposed the infrared source \astrobj{IRAS 18357-0604} to be an analogue of IRC\,+10420. The present paper is devoted to IRC\,+10420, certainly the member of this scarce category that is most studied at different spatial scales.

\begin{figure*}[ht]
\begin{center}
\includegraphics[width=125mm]{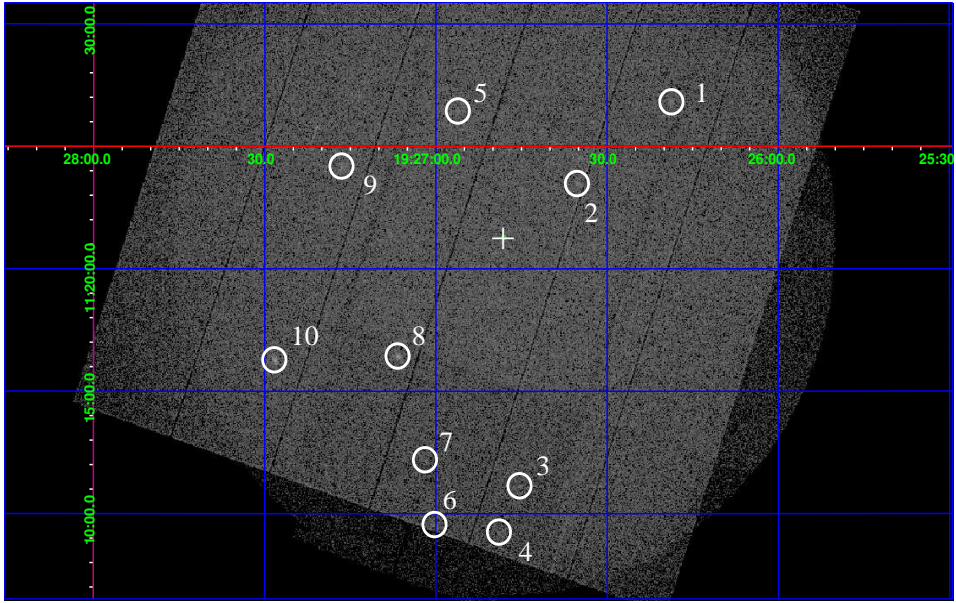}
\caption{Composite image of the EPIC field of view centered at the position of IRC\,+10420 (indicated by the + sign). Detected X-ray point sources are marked by white circles, and associated numbers refer to Table\,\ref{detect}. \label{epicimage}}
\end{center}
\end{figure*}

According to the study of \citet{oudmaijer1996}, the spectral energy distribution of IRC\,+10420 has changed significantly over several decades, simultaneously with an increase of effective temperature of about 1000\,K. This object is surrounded by shells of material related to the previous red supergiant episode which gave rise to substantial mass-loss \citep{kastner1995,humphreys1997}. The mass loss rate determined by \citet{oudmaijer1996} is of the order of 5\,$\times$\,10$^{-4}$\,M$_\odot$\,yr$^{-1}$ assuming IRC\,+10420 is at a distance of 3.5\,kpc. In addition, the infrared emission excess of IRC\,+10420 reveals the presence of a hot dust component organized in a circumstellar disk/annulus. According to speckle interferometry \citep{blocker1999}, the circumstellar shell can be interpreted in terms of an enhanced mass-loss phase roughly 60 to 90\,yr ago. The dust shell contributes to about 40\,$\%$ of the total flux at 2.11\,$\mu$m, with the remaining 60\,$\%$ being attributed to the central object. The same authors derived a bolometric luminosity of the order of 3\,$\times$\,10$^5$\,L$_\odot$ or 6\,$\times$\,10$^5$\,L$_\odot$ for distances of 3.5 or 5\,kpc respectively, corresponding to the range of distances generally adopted for this object. More recently, interferometric observations with the VLTI \citep{dewit2008} probed the vicinity of the central star, and confirmed the lack of significant amount of dust very close (at the milli-arcsec scale) to the star, providing support for a dust shell located further out, at an angular separation of about 70\,mas from the star. 

So far, IRC\,+10420 has never been a target for modern X-ray observatories. As a massive star with outflowing material, one may expect similar processes as those active in stellar winds of massive stars (OB or WR) to be at work in the envelope of such objects. In particular, local rise of plasma temperature as a result of interactions between shells moving out at different speeds is likely to produce thermal X-rays \citep{FeldX}, with a spectral hardness depending on the pre-shock velocities of colliding shells. Following this idea, X-ray astrophysics may be a valuable tool to derive information on IRC\,+10420 in complementarity with other techniques previously applied to this object. This paper reports on the results of the very first dedicated observation in X-rays of IRC\,+10420 and its surroundings, using the {\it XMM-Newton} satellite. The data reduction procedure is described in Sect.\,\ref{observ}. Our results and discussion related to IRC\,+10420 are provided in Sect.\,\ref{results}. The discussion of other X-ray point sources detected in the field of view is presented in Sect.\,\ref{sources}. We finally conclude in Sect.\,\ref{concl}.

\section{Observation and data processing} \label{observ}

\subsection{Available data}
IRC\,+10420 has been a target of the {\it XMM-Newton} satellite \citep{xmm} during the 10th Announcement of Opportunity (AO10, rev.\,2180), on 3rd November, 2011 (Julian Date 2,455,896.375), under proposal ID\,067107 (PI: M. De Becker). EPIC instruments \citep{mos,pn} were operated in Full Frame mode, and the medium filter was used to reject optical light. The exposure times were 12.7 and 11.0\,ks for MOS and pn instruments, respectively. The exposure was affected by a rather high background level due to a soft proton flare. However, as most of the short exposure was contaminated, we refrained from rejecting affected time intervals. Data were processed using the {\it XMM-Newton} Science Analysis Software (SAS) v.12.0.0 on the basis of the Observation Data Files (ODF) provided by the European Space Agency (ESA). Event lists were filtered using standard screening criteria (pattern $\leq$\,12 for MOS and pattern $\leq$\,4 for pn).

The Optical/UV Monitor \citep[OM,][]{om} was also operating simultaneously with the EPIC observations. The field surrounding IRC\,+10420 was observed in imaging mode successively in the UVM2 and UVW2 bands (2050–-2450 and 1800–-2250\AA\,, respectively), with exposure times of 4.4\,ks. We obtained images and source lists through the {\it omichain} SAS meta-task. 

\subsection{Source detection}
In X-rays, the first data products we constructed from EPIC event lists were images. The most obvious information obtained from these images is the apparent lack of detection of IRC\,+10420, located very very close to the center of the field of view. In addition, the visual inspection of images revealed the presence of a few point sources in the field of view.

We created images from MOS and pn event lists using various energy filters, corresponding to energy bands of 0.3--1.0\,keV ({\it Soft} band, referred to as S hereafter), 1.0--2.0\,keV ({\it Medium} band, M) and 2.0--10.0\,keV ({\it Hard} band, H). We applied the standard procedure for source detection in EPIC data using the {\it edetect\_chain} meta-task available within the SAS. We used a logarithmic likelihood for the detection of sources equal to 12 (therefore slightly more conservative than the default value of 10). The source detection procedure was executed in all energy bands, either separately on each EPIC data set or simultaneously for the three instruments. The procedure led to the detection of a few sources for EPIC-MOS data, with no detection coincident with the coordinates of IRC\,+10420. The same procedure applied to the EPIC-pn data set did not allow the detection of IRC\,+10420, but led to the detection of many spurious sources in addition to the sources already revealed by eye inspection of the images. The spurious detections were mostly located at the boundaries of the pn CCDs, along with a few ones suspiciously aligned along the same pixel rows. In total, we report on the detection of 10 X-ray point sources in the field of view.  A composite EPIC image with over-plotted circular regions centered on the position of the detected sources is given in Fig.\,\ref{epicimage}.\\

In the UV, we report on the detection of 7 sources in the UVM2 band, and of 7 sources in the UVW2 band, with 6 sources in common between the two spectral domains. The most puzzling feature in OM data concerns the apparent detection of a significant point source at the position of IRC\,+10420 in the UVW2 band, but we caution that it should be attributed to leakage of red photons that is especially problematic for this filter\footnote{See the OM calibration status document, XMM-SOC-CAL-TN-0019 Issue 6.0 (January 2011), available at http://xmm2.esac.esa.int/docs/documents/CAL-TN-0019.pdf.}. It appears indeed that bright sources in the red domain can lead to a substantial fake detection, especially in the UVW2 filter. The other UV sources (a priori real sources) are not counterparts of the X-ray sources detected in the EPIC field of view.

\section{IRC\,+10420}\label{results}

\subsection{Upper limits on the X-ray flux} \label{upper}

Even though our main target is not detected, we could derive upper limits on the X-ray flux at the expected position of IRC\,+10420. As a first approach, we built sensitivity maps using the SAS task {\it esensmap} for the three EPIC data sets, for a logarithmic likelihood of 12. The sensitivity maps were calculated in a large energy band, i.e. 0.3--10\,keV. We then considered the sensitivity value measured at the position of our target as a first guess of the upper limit on the X-ray count rate: 0.0034, 0.0031 and 0.0080 cnt\,s$^{-1}$for MOS1, MOS2 and pn, respectively. The aim-point of the observation was set to the position of IRC\,+10420, that is therefore located in the part of the EPIC field of view where the exposure map reaches its highest value. 

\begin{figure}[ht]
\begin{center}
\includegraphics[width=85mm]{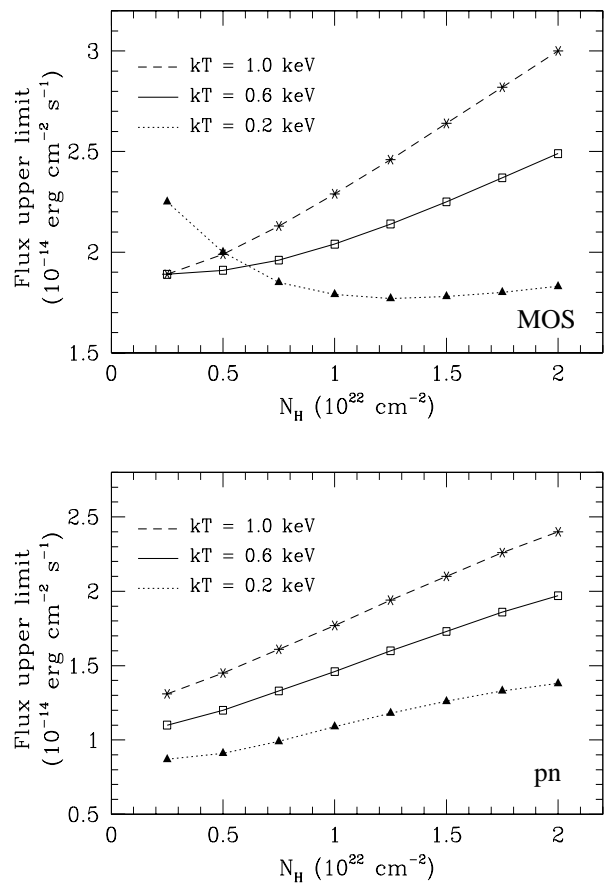}
\caption{Upper limits derived on the X-ray flux between 0.3 and 10.0\,keV for EPIC-MOS ({\it top}) and EPIC-pn ({\it bottom}) at the position of IRC\,+10420, as a function of assumed values for the plasma temperature and the absorbing column. \label{ulpimms}}
\end{center}
\end{figure}

As a second approach, we extracted spectra in circular regions (radius of 30'') centered at the coordinates of IRC\,+10420 (RA:\,19$^h$\,26$^m$\,48.10$^s$; DEC:\,+11$^\circ$\,21'\,16.74''; J2000), and at four other positions devoid of any point source according to our detection procedure, and located within about 3 arcminutes from the center of the field of view. Response and ancillary matrices were computed using the dedicated SAS tasks. We then measured the count rates on the spectra using the XSPEC\footnote{http://heasarc.gsfc.nasa.gov/docs/xanadu/xspec/, see also \citet{xspec1996}.} software (v.12.5.1). We repeated the procedure for the three EPIC instruments. The count rate values measured at the five selected locations on the image are very similar, generally within 1-$\sigma$. These measurements lend therefore support to the idea that the X-ray emission level at the position of IRC\,+10420 is consistent with that of the background, in agreement with its non-detection. To derive an upper limit on the count rate of IRC\,+10420, we extracted events in a smaller circular region (radius of 6''), corresponding to the peak of the Point Spread Function (PSF) of any expected X-ray source at a given position. Such a narrow extraction region is motivated by the need to estimate any excess in the X-ray emission with respect to the surrounding background level. The uncorrected count numbers (C) are given in Table\,\ref{counts}. We selected a count threshold (C$_\mathrm{max}$) corresponding to a logarithmic likelihood of 12, which translates into a probability to find a count number in excess of the critical value of about 6\,$\times$\,10$^{-6}$, under the null hypothesis of pure background fluctuations. The count excess, i.e. C$_\mathrm{max}$ $-$ C, was then divided by the effective exposure time to derive an upper limit on the count rate (associated to the extraction radius). The numbers were finally corrected for the encircled energy fraction of about 40\,$\%$ (for a 6'' radius, see the XMM User's Handbook) to determine the upper limits on the count rate (CR) of IRC\,+10420, for the three EPIC instruments (CR$_\mathrm{cor}$\,=\,CR/0.4). The same procedure was applied by \citet{gossetwr40} to the case of the non detection of WR\,40. All the results are given in Table\,\ref{counts}. These upper limits are consistent with those derived from the sensitivity maps, and will be considered to be our measured upper limits on the count rates of IRC\,+10420.

\begin{table*}[ht]
\caption{Estimates of the upper limits at the position of IRC\,+10420 on the count rate for all EPIC instruments.\label{counts}}
\begin{center}
\begin{tabular}{l c c c c c c}
\hline
 & C & C$_\mathrm{max}$ & C$_\mathrm{max}$ $-$ C & Eff. exp. time & CR & CR$_\mathrm{cor}$  \\
 & (cnt) & (cnt) & (cnt) & (s) & (cnt\,s$^{-1}$) & (cnt\,s$^{-1}$) \\
\hline
\vspace*{-0.2cm}\\
MOS1 & 11 & 28.7 & 17.7 & 12470 &1.4\,$\times$\,10$^{-3}$ & 3.5\,$\times$\,10$^{-3}$ \\
MOS2 & 8 & 23.6 & 15.6 & 12480 & 1.3\,$\times$\,10$^{-3}$ & 3.3\,$\times$\,10$^{-3}$ \\
pn & 44 & 76.4 & 32.4 & 9600 & 3.4\,$\times$\,10$^{-3}$ & 8.5\,$\times$\,10$^{-3}$ \\
\vspace*{-0.2cm}\\
\hline
\end{tabular}
\end{center}
\end{table*}

We then used the WEBPIMMS\footnote{http://ledas-www.star.le.ac.uk/pimms/w3p/w3pimms.html} tool to convert these count rates into flux upper limits, assuming an absorbed APEC optically thin thermal emission model (using solar abundances). As we don't have any a priori idea of the most adequate physical parameters, we assumed different values for the absorbing hydrogen column (N$_\mathrm{H}$, expressed in 10$^{22}$\,cm$^{-2}$) and for the plasma temperature (kT, expressed in keV). Our model grid covers N$_\mathrm{H}$ values between 0.25 and 2.0\,$\times$\,10$^{22}$\,cm$^{-2}$, and temperatures between 0.2 and 1.0\,keV (i.e. between 2\,$\times$\,10$^{6}$ and 1.2\,$\times$\,10$^{7}$\,K). The upper limits on the flux, expressed in erg\,cm$^{-2}$\,s$^{-1}$, are plotted in Fig.\,\ref{ulpimms}. The general trend followed by our flux estimates as a function of model parameters is an increase of the flux as the absorbing column increases, for a given temperature. This translates the fact that for higher N$_\mathrm{H}$ values, only the higher energy photons are significantly detected, and consequently a given count rate converts into a higher energy flux. This effect is obviously more pronounced as higher plasma temperatures are considered, explaining why the highest fluxes are obtained for the highest kT values. The only exception to this trend is seen for MOS fluxes with kT\,=\,0.2\,keV. Such a low temperature would produce a very soft X-ray spectrum, with most photons emitted at energies where the photoelectric absorption is efficient, up to a level where the X-ray flux becomes insensitive to absorption, producing the almost flat dotted curve in the upper panel of Fig.\,\ref{ulpimms}. The fact that such a behavior is not observed for pn fluxes comes from the significantly different instrumental responses of MOS and pn detectors at very low X-ray energies. Globally, depending on the instrument and on the model, we derived typical upper limits of the order of 1--3\,$\times$\,10$^{-14}$\,erg\,cm$^{-2}$\,s$^{-1}$. We emphasize that this value stands for the X-ray flux absorbed by both the circumstellar and interstellar media, and not for the un-attenuated one.  

\subsection{Discussion}\label{disc}

For the purpose of this discussion, one should briefly comment on the scenarios likely to produce X-rays in such an evolved massive star:
\begin{enumerate}
\item[-] {\it Embedded shocks}. The strong stellar wind is likely to undergo instabilities leading to intrinsic shocks, heating the shocked plasma up to temperatures of several 10$^6$\,K \citep[e.g.][]{FeldX}. The typical temperature of the X-ray emitting plasma is dependent on the pre-shock velocity of out-moving and colliding shells, with values ranging between 1.4\,$\times$\,10$^{5}$ to 3.4\,$\times$\,10$^{6}$\,K, respectively for pre-shock velocities of 100 and 500\,km\,s$^{-1}$. The lower values in this range will lead to quite soft thermal X-ray spectra, with the bulk of their flux produced at energies where the circumstellar absorption is the most efficient. This might be the main reason for the non-detection of IRC\,+10420 in X-rays.
\item[-] {\it Chromospheric-like activity}. In such a scenario, the plasma temperature can reach somewhat higher values, up to about 10$^{7}$\,K \citep{gudel2004}. Such temperatures are expected to lead to harder X-ray spectra, therefore more weakly affected by photoelectric absorption due to the dense circumstellar material. However, \citet{clark2008} caution that it is not clear that yellow hypergiants evolving from rather massive stars will develop such a chromospheric activity.
\end{enumerate} 

In the absence of detection, one could only speculate on the origin of this lack of apparent X-ray activity. Either the source is intrinsically very weak, or it is significantly absorbed by shells of material surrounding the potentially emitting region. According to \citet{oudmaijer1996}, one may expect the outflowing material in IRC\,+10420 to be quite abundant, and whatever the intrinsic X-ray emission level, it should be significantly attenuated by these large amounts of material. In such circumstances, we may consider to make the following assumption: as the X-ray flux emerging from the source is most probably substantially affected by local photoelectric absorption, the subsequent impact of interstellar absorption should be quite weak, and consequently negligible. This is what one may call the {\it strong circumstellar absorption hypothesis}. According to this assumption, fluxes derived above could be considered as emerging from the circumstellar environment of the star, and be confronted to bolometric fluxes of the target. \citet{blocker1999} estimate the bolometric flux of IRC\,+10420, assuming a distance of 3.5\,kpc, to be 8.5\,$\times$\,10$^{-7}$\,erg\,cm$^{-2}$\,s$^{-1}$, leading to an upper limit on the f$_\mathrm{X}$/f$_\mathrm{bol}$ ratio of 1.1--3.5\,$\times$\,10$^{-8}$. Assuming a distance of about 5\,kpc, the bolometric flux should be a factor 2 larger, leading to a f$_\mathrm{X}$/f$_\mathrm{bol}$ ratio of 0.5--1.8\,$\times$\,10$^{-8}$. Alternatively, if the so-called strong circumstellar absorption hypothesis is not valid, one can only speculate on the respective interstellar and circumstellar contributions to the total absorption, and upperlimits on the f$_\mathrm{X}$/f$_\mathrm{bol}$ ratio corrected for ISM absorption could not be determined.

In an attempt to estimate the interstellar contribution to the X-ray absorption, we considered a E(B\,--\,V) color excess of 1.84 determined by \citet{humphreys2002} on the basis of equivalent width measurements of Diffuse Interstellar Bands in the optical spectrum of IRC\,+10420. According to the dust-to-gas relation given by \citet{Boh}, we derived an interstellar N$_\mathrm{H}$\,$\sim$\,1.1\,$\times$\,10$^{22}$\,cm$^{-2}$. Such a large value is operating in addition to the circumstellar material, and should therefore be considered as a lower limit on the total absorbing column. This suggests that only values plotted on the right parts of Fig.\,\ref{ulpimms} may be relevant.

In the context of X-ray emission from single massive stars, one could consider the so-called canonical f$_\mathrm{X}$/f$_\mathrm{bol}$ ratio of the order of 10$^{-7}$ for regular O-type stars \citep{sanalxlbol,owocki2013}. Our derived upper limits on the flux ratio provide evidence that the X-ray emission from IRC\,+10420 is not higher than that of regular O-type stars, and even probably somewhat lower. For instance, the X-ray investigation of two extreme O-type supergiants on the way to the WN stage revealed flux ratios slightly lower than 10$^{-7}$, in agreement with the expected enhanced absorption due to their dense stellar winds \citep{debeckerofplusxmm}, as discussed also by \citet{owocki2013}. In the case of IRC\,+10420, a much larger circumstellar absorption is expected, and significantly lower flux ratios can be anticipated. On the other hand, our upper limits are compatible with the 10$^{-8}$--10$^{-6}$ range reported by \citet{ayres2005} for the chromospheric X-ray emission process from later type supergiants. The main conclusion that can be drawn from these f$_\mathrm{X}$/f$_\mathrm{bol}$ values is that the sensitivity of our short XMM-Newton exposure is sufficient to provide strong constraints on the X-ray emission mechanism putatively operating in IRC\,+10420. 

\begin{table*}[ht]
\caption{X-ray sources detected in EPIC data, including their count rates and hardness ratios.\label{detect}}
\begin{center}
\begin{tabular}{c c c c c c c c}
\hline
\vspace*{-0.2cm}\\
XID & Name & \multicolumn{3}{c}{Count rates (10$^{-2}$\,cnt/s)} & & \multicolumn{2}{c}{Hardness ratios} \\
\cline{3-5}\cline{7-8}
\vspace*{-0.2cm}\\
 &  & MOS1 & MOS2 & pn &  & 1 & 2 \\
\hline
1 & XMMU\,J192618.7+112649 & -- & 0.65\,$\pm$\,0.17 & 1.91\,$\pm$\,0.39 &  & --0.11\,$\pm$\,0.19 & 0.20\,$\pm$\,0.24 \\
2 & XMMU\,J192635.3+112329 & 0.46\,$\pm$\,0.11 & 0.41\,$\pm$\,0.10 & 0.93\,$\pm$\,0.23 &  & 1.00\,$\pm$\,0.34 & 0.34\,$\pm$\,0.20 \\
3 & XMMU\,J192645.4+111108 & -- & -- & 1.66\,$\pm$\,0.35 &  & 1.00\,$\pm$\,0.14 & 0.46\,$\pm$\,0.17 \\
4 & XMMU\,J192650.0+110915 & -- & -- & 1.84\,$\pm$\,0.45 &  & 0.80\,$\pm$\,0.37 & 0.51\,$\pm$\,0.17 \\
5 & XMMU\,J192656.2+112626 & 0.32\,$\pm$\,0.09 & 0.42\,$\pm$\,0.11 & 1.35\,$\pm$\,0.23 &  & --0.39\,$\pm$\,0.15 & --1.00\,$\pm$\,0.50 \\
6 & XMMU\,J192700.2+110933 & -- & 0.57\,$\pm$\,0.14 & 3.32\,$\pm$\,0.52 &  & --0.41\,$\pm$\,0.13 & --0.14\,$\pm$\,0.33 \\
7 & XMMU\,J192701.9+111212 & -- & -- & 1.42\,$\pm$\,0.36 &  & --0.05\,$\pm$\,0.22 & --0.05\,$\pm$\,0.38 \\
8 & XMMU\,J192706.7+111626 & -- & 1.72\,$\pm$\,0.18 & 4.88\,$\pm$\,0.41 &  & 0.81\,$\pm$\,0.09 & 0.27\,$\pm$\,0.08 \\
9 & XMMU\,J192716.6+112411 & 0.93\,$\pm$\,0.18 & 0.68\,$\pm$\,0.15 & 1.97\,$\pm$\,0.38 &  & 0.06\,$\pm$\,0.16 & --0.24\,$\pm$\,0.30 \\
10 & XMMU\,J192728.3+111617 & 2.26\,$\pm$\,0.29 & 2.47\,$\pm$\,0.29 & 7.64\,$\pm$\,0.63 &  & 0.23\,$\pm$\,0.08 & --0.03\,$\pm$\,0.10 \\
\vspace*{-0.2cm}\\
\hline
\end{tabular}
\end{center}
\end{table*}

\subsection{Spectral energy distribution}

We first considered the upper limits in X-rays derived in Section\,\ref{upper}. We converted the flux upper limit into a flux density assumed to be constant in the 0.3--10.0\,keV bandpass, and we obtained F$_\lambda$\,=\,2.5--7.5\,$\times$\,10$^{-16}$\,\,erg\,cm$^{-2}$\,s$^{-1}$\,\AA\,$^{-1}$. In the ultraviolet, the apparently detected count rate in the UVW2 band is (4.818\,$\pm$\,1.249)\,$\times$\,10$^{-2}$ cnt\,s$^{-1}$. We used the rate-to-flux conversion factor adequate for the UVW2 energy band, i.e. 5.7\,$\times$\,10$^{-15}$\,\,erg\,cm$^{-2}$\,cnt$^{-1}$\,\AA\,$^{-1}$ (see the {\it XMM-Newton Optical and UV Monitor (OM) Calibration Status} document, Talavera 2011\footnote{http://xmm2.esac.esa.int/docs/documents/CAL-TN-0019.pdf}), and we converted this count rate into a flux density of (2.75\,$\pm$\,0.71)\,$\times$\,10$^{-16}$\,\,erg\,cm$^{-2}$\,s$^{-1}$\,\AA\,$^{-1}$. In the following, this value will be considered as an upper limit on the flux density between 1800 and 2250\,\AA\,, as the apparent detection is very unlikely. We also considered the count rate of the faintest source detected in the UVM2 band, i.e. (4.504\,$\pm$\,1.356)\,$\times$\,10$^{-2}$\,cnt\,s$^{-1}$, as an upper limit for IRC\,+10420. With a rate-to-flux conversion factor of 2.2\,$\times$\,10$^{-15}$\,\,erg\,cm$^{-2}$\,cnt$^{-1}$\,\AA\,$^{-1}$, we obtained an upper limit on the flux density of (9.91\,$\pm$\,3.00)\,$\times$\,10$^{-17}$\,\,erg\,cm$^{-2}$\,s$^{-1}$\,\AA\,$^{-1}$. We finally derived flux densities from the infrared and visible magnitudes quoted in Table\,\ref{counterparts} for the counterparts of IRC\,+10420, using the magnitude-to-flux density converter hosted by the Spitzer Science Center\footnote{http://ssc.spitzer.caltech.edu/warmmission/propkit/pet/magtojy/index.html}.

\begin{figure}[ht]
\begin{center}
\includegraphics[width=85mm]{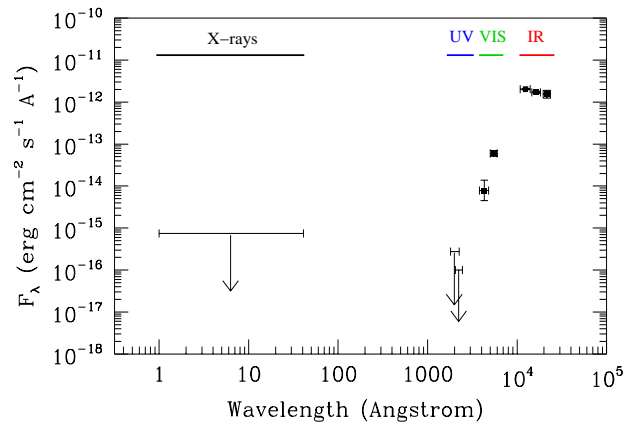}
\caption{Observational SED of IRC\,+10420 from X-rays to the infrared domain. The downward arrows stand for upper limits derived in the X-ray and UV domains. \label{sed}}
\end{center}
\end{figure}

\begin{figure}[ht]
\begin{center}
\includegraphics[width=85mm]{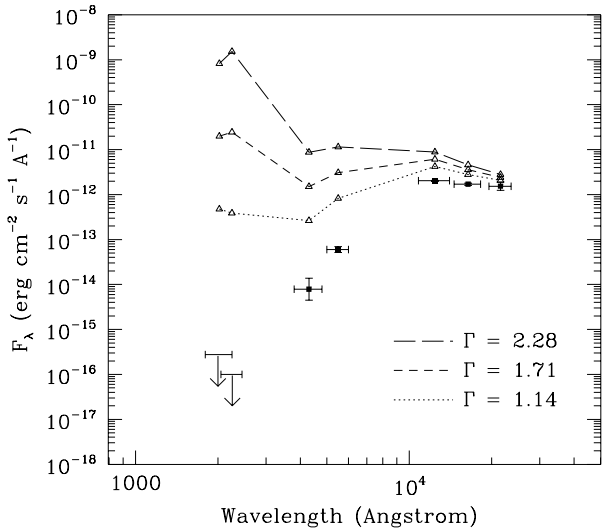}
\caption{Corrected SED assuming three different extinction regimes, from the UV to the infrared domain. The data points for a given extinction have been connected using long-dashed, short-dashed and dotted lines, respectively, for the sake of clarity. \label{sedcor}}
\end{center}
\end{figure}

The observational Spectral Energy Distribution (SED) is plotted in Fig.\,\ref{sed}. The upper limits derived in X-rays (EPIC) and in the UV domain (OM) are illustrated by downward arrows. The width of each data point stands for the wavelength domain of each measurement. The SED shows clearly that the decrease in the flux density extends from the infrared to the ultraviolet. The fact that IRC\,+10420 is not detected in the UV bands agrees with the idea that IRC\,+10420 is a highly obscured object, with a strong absorption close to the dust extinction peak at about 2200\,\AA\,. 

As stated above, its putative detection at a shorter wavelength (in the UVW2 band) is surprising and should be attributed to a red leak of the filter in the presence of a bright red source. As an additional verification, we corrected the observed SED for the extinction. To do so, we used the parametric determination of $t_\lambda = A_\lambda/A_V$ given by \citet{extinctioncurve}, and we determined corrected flux densities according to the following relation:
$$ \mathrm{F}_{\lambda,corr} = \mathrm{F}_{\lambda}\,10^{\Gamma\,t_\lambda}$$
with $\Gamma\,=\,0.4\,R_V\,E(B-V)$. Assuming $R_V$\,=\,3.1 and $E(B-V)$\,=\,1.84 \citep{humphreys2002}, we obtain $\Gamma$\,$\sim$\,2.28. According to the parametric relations given by \citet{extinctioncurve}, we derive $t_{2000\,\AA\,}$ = 2.84 and $t_{2200\,\AA\,}$ = 3.15. In such circumstances, the corrected flux density in the UVW2 band increases to a level that is indeed below the upper limit derived in the UVM2 band. However, the same correction applied to the visible and near-infrared flux densities leads to another unexpected result: the corrected UV flux densities are about two orders of magnitude higher than in the visible (see Figure\,\ref{sedcor}) which is unrealistic.

We also corrected the measured flux densities assuming weaker extinctions, respectively for $\Gamma$ values equal to 0.75 and 0.5 times the value based on \citet{humphreys2002}. We have to consider an extinction that is two times weaker than a priori expected to shift the UV flux densities to a level similar to that obtained in the visible domain, which is in contradiction with the expected high value of the $\mathrm{E(B-V)}$ color excess toward IRC\,+10420. As a result, the SED corrected for extinction demonstrates that, if the UVW2 source was real, it would point to an unexpected UV excess very difficult to explain on a physical basis. This lends significant support to the idea that the putative UV source detected in the UVW2 band is a fake, whose existence is explained by the strong leak of red photons in this filter.

\section{Detected point sources}\label{sources}

\subsection{X-ray sources}

The census of detections is summarized in Table\,\ref{detect}.  We report on the detection of 10 sources with identifiers attributed according to the standard naming conventions for {\it XMM-Newton} sources. On the basis of the count rates measured in the three energy bands, we calculated hardness ratios defined by these relations: $HR_1 = (M - S)/(M + S)$ and $HR_2 = (H - M)/(H + M)$, where $S$, $M$ and $H$ stand for the count rates measured in the S, M, and H bands, respectively. The quoted hardness ratios were determined using EPIC-pn count rates. Our values point to some especially soft sources (XID5, XID6 and XID7), but the influence of the soft proton flares that affected our EPIC exposure severely limits our capability to constrain properly the X-ray properties of these sources.

\begin{table*}[ht]
\caption{Sources detected in OM data, including their UV magnitudes. Source number 4 corresponds to IRC\,+10420.\label{omsources}}
\begin{center}
\begin{tabular}{c c c c c c}
\hline
UVID & $\alpha$\,[J2000] & $\delta$\,[J2000] & UVM2 & UVW2 & Comment  \\
\hline
\vspace*{-0.2cm}\\
1 & 19 26 43.40 & +11 18 39.9 & 15.669\,$\pm$\,0.029 & 15.877\,$\pm$\,0.052 & -- \\
2 & 19 26 44.60 & +11 19 05.1 & 16.036\,$\pm$\,0.035 & 16.350\,$\pm$\,0.071 & -- \\
3 & 19 26 45.01 & +11 17 35.9 & 12.000\,$\pm$\,0.001 & 12.092\,$\pm$\,0.001 & -- \\
4 & 19 26 48.10 & +11 21 16.7 & -- & 18.159\,$\pm$\,0.282 & false detection \\
5 & 19 26 53.34 & +11 23 15.8 & 15.466\,$\pm$\,0.026 & 15.868\,$\pm$\,0.052 & -- \\
6 & 19 26 53.65 & +11 25 41.3 & 17.758\,$\pm$\,0.109 & 18.073\,$\pm$\,0.269 & -- \\
7 & 19 26 59.98 & +11 21 33.2 & 17.239\,$\pm$\,0.075 & 17.427\,$\pm$\,0.158 & -- \\
8 & 19 27 05.11 & +11 18 43.2 & 19.138\,$\pm$\,0.329 & -- & -- \\
\vspace*{-0.2cm}\\
\hline
\end{tabular}
\end{center}
\end{table*}

\begin{table*}[ht]
\caption{Potential counterparts of the X-ray and UV sources, including their magnitudes in several bands. We caution that UVID4(*) is a false detection, and the information about IRC\,+10420 is given as it was used to build the SED. \label{counterparts}}
\begin{center}
\begin{tabular}{c c c c c c c c c c c c}
\hline
\vspace*{-0.2cm}\\
ID & \multicolumn{5}{c}{2MASS} & & \multicolumn{5}{c}{GSC\,2.2} \\
\cline{2-6}\cline{8-12}
\vspace*{-0.2cm}\\
 & d\,('') & ID & J & H & K$_S$ &  & d\,('') & ID & B & V & R \\
\hline
XID1 & 3.9 & 19261893+1126508 & 12.95 & 12.39 & -- &  & 3.9 & N0232132179292 & 16.08 & -- & 14.14 \\
XID2 & 4.3 & 19263537+1123331 & 15.94 & 14.88 & 14.70 &  & -- & -- & -- & -- & -- \\
XID3 & -- & -- & -- & -- & -- & & -- & -- & -- & -- & -- \\
XID4 & 3.0 & 19264991+1109123 & 14.45 & 13.29 & 12.96 &  & -- & -- & -- & -- & -- \\
XID5 & 2.9 & 19265606+1126282 & 10.46 & 10.26 & 10.13 &  & 2.9 & N0232132526 & 12.11 & 11.45 & -- \\
XID6 & 3.4 & 19270030+1109360 & 12.18 & 11.73 & 11.58 &  & 3.6 & N0232132162346 & 14.76 & -- & 13.38 \\
XID7 & 1.0 & 19270188+1112129 & 13.39 & 12.96 & 12.82 &  & 1.1 & N0232132165565 & 16.52 & -- & 14.58 \\
XID8 & -- & -- & -- & -- & -- &  & -- & -- & -- & -- \\
XID9 & 1.6 & 19271649+1124109 & 12.31 & 11.55 & 11.29 &  & 0.9 & N0232101160824 & 16.32 & -- & 14.14 \\
XID10 & 2.0 & 19272816+1116175 & 11.26 & 10.62 & 10.46 &  & 2.1 & N0232132170442 & 14.84 & -- & 12.95 \\
\hline
UVID1 & 0.3 & 19264341+1118396 & 11.63 & 11.46 & 11.37 &  & 0.6 & N0232132550 & 13.20 & -- & 12.12 \\
UVID2 & 0.4 & 19264461+1119047 & 11.06 & 10.84 & 10.76 &  & 0.2 & N0232132548 & 13.10 & -- & 12.05 \\
UVID3 & 0.7 & 19264502+1117352 & 9.43 & 9.37 & 9.32 &  & 0.6 & N0232132557 & 10.38 & 10.66 & -- \\
UVID4(*) & 0.1 & 19264809+1121167 & 5.47 & 4.54 & 3.61 &  & 0.3 & N0232132545	 & 14.83 & 11.99 & -- \\
UVID5 & 0.3 & 19265335+1123157 & 10.59 & 10.34 & 10.25 &  & 0.1 & N0232132536 & 12.73 & 11.79 & -- \\
UVID6 & 2.2 & 19265356+1125431 & 11.61 & 11.37 & 11.27 &  & 2.4 & N0232132528 & 13.50 & -- & 12.59 \\
UVID7 & 0.5 & 19270000+1121328 & 11.55 & 11.23 & 11.10 &  & 0.6 & N0232132543 & 13.90 & -- & 12.64 \\
UVID8 & 0.4 & 19270513+1118432 & 12.69 & 12.36 & 12.23 &  & 0.6 & N0232132173296 & 15.08 & -- & 13.76 \\
\vspace*{-0.2cm}\\
\hline
\end{tabular}
\end{center}
\end{table*}

We extracted spectra for the brightest objects in every data sets where they were detected, using circular spatial filters (radius of 30'') centered on the sources. We also extracted background spectra in annular regions centered on the same location as the source region. Spectra were grouped to get at least 5 counts per energy bin. We also computed response and ancillary files using the dedicated SAS tasks for each source. Even in the cases of the two brightest sources, i.e XID8 and XID10, the quality of the spectra is very poor. Apart from the fact that the exposure was rather short and that the sources are faint, the main reason for this low quality is the high background level that affects the whole exposure. It has been previously explained that data affected by such soft proton flares should only be used if the studied sources are bright enough, otherwise the background correction could not be applied properly \citep[see e.g.][]{DeB159176}. This effect is obvious in the spectra with data points presenting huge error bars, mostly in the harder part of the spectrum where the source/background ratio is quite low. As we could not filter the event lists without rejecting most of the exposure to get rid of this issue, there is no way to clean the spectra obtained with the present data set. For these reasons, we limit our spectral analysis to XID10, and we emphasize that the physical parameters we derived should be considered with caution. The results of the simultaneous fit of a wabs*apec model to EPIC data yield a plasma temperature of 1.00\,$\pm$\,0.13\,keV with N$_\mathrm{H}$\,=\,(1.37\,$\pm$\,0.17)\,$\times$\,10$^{22}$\,cm$^{-2}$ ($\chi^2_\nu$ \,=\,1.02, for 334 degrees of freedom). On the basis of this model, the estimated flux is $\sim$\,10$^{-13}$\,erg\,cm$^{-2}$\,s$^{-1}$, but with a large error considering the large relative error on the normalization parameter of the best-fit model. The plasma temperature is compatible with the hardness ratios derived for the same source (see Table\,\ref{detect}).

\subsection{Counterparts at other wavelengths}

The census of UV sources detected with the Optical Monitor is given in Table\,\ref{omsources}. We emphasize that none of these sources are counterparts of the X-ray sources quoted in Table\,\ref{detect}. Most UV sources are detected in the UVM2 and UVW2 bands, except UVID4 (i.e. false detection of IRC\,+10420) present only in the latter and UVID8 detected only in the former band. 

We cross-correlated the position of these X-ray sources with the Two-Micron All Sky Survey (2MASS) catalogue \citep{2mass} in the infrared, and with the Guide Star Catalogue (GSC) in the visible domain (STScI). The census of potential infrared and visible counterparts is given in Table\,\ref{counterparts}, where the angular distance (d) between the point sources and their counterparts is specified in each case. We arbitrarily set the maximum correlation radius to 5\,arcseconds. All UV sources have counterparts in the infrared and in the visible. However, we did not identify any counterpart for XID3 and XID8, even though the latter is the second brightest X-ray source in the EPIC field of view (at least at the time of the XMM-Newton observation).

\section{Conclusions}\label{concl}

We report on the first dedicated observation of the yellow hypergiant IRC\,+10420 at short wavelengths, i.e. in soft X-rays and in the ultraviolet, using the XMM-Newton satellite. The target is not detected in X-rays, and we derived conservative upper limits on the X-ray flux of 1--3\,$\times$\,10$^{-14}$\,erg\,cm$^{-2}$\,s$^{-1}$ (between 0.3 and 10.0\,keV). The lack of detection in X-rays is either due to a weak intrinsic emission (if any), or to a strong absorption notably by circumstellar material proved by previous studies to be quite abundant. Assuming that the circumstellar absorption is very strong, therefore implying a moderate impact of subsequent interstellar absorption, we converted the flux upper limit into an upper limit on the f$_\mathrm{X}$/f$_\mathrm{bol}$ ratio (corrected for ISM absorption) of 1.1--3.5\,$\times$\,10$^{-8}$, that is not enough constraining to be confronted to the expected flux ratio of regular massive stars ($\sim$\,10$^{-7}$), even though it likely points to a somewhat lower intrinsic emission. We also derived the first upper limits on the ultraviolet emission from IRC\,+10420, between 1800 and 2250\,\AA\, and between 2050 and 2450\,\AA\,. Using our new measurements, we derived an observational spectral energy distribution ranging from X-rays to the infrared. The wide band spectral energy distribution is qualitatively explained in the context of an object strongly obscured due to optically thick circumstellar material, leading to a steeply decreasing flux density from the infrared up to the ultraviolet. It turns out that any detection of IRC\,+10420 is unlikely with present X-ray facilities, and that much larger collecting areas as those ones envisaged for future generations of observatories are mandatory (but not necessarily sufficient) to clarify the nature of the putative X-ray activity related to IRC\,+10420, and potentially to other yellow hypergiants with circumstellar shells such as HD\,179821, Hen\,3-1379 and IRAS\,18357-0604. 

We also investigated the point sources present in the field of view of the X-ray and UV detectors on-board XMM-Newton. We report on the detection of 10 X-ray and 8 UV sources (including the false detection of IRC\,+10420), without positional coincidence between these two sets of point sources. X-ray hardness ratios were determined, but the contamination of the exposure by a high background event prevented us to derive detailed properties of these X-ray point sources. We cross-correlated the positions of these sources with infrared and visible catalogues, and we found probable counterparts for all but two X-ray sources, and for all UV sources.

\section*{Acknowledgements}
This research is supported in part through the PRODEX XMM/Integral contract. The authors warmly thank the XMM panel members for supporting the present project. The XMM-Newton Science Operations Center (SOC) is kindly acknowledged for the scheduling of the observation. This publication makes use of data products from the Two Micron All Sky Survey, which is a joint project of the University of Massachusetts and the Infrared Processing and Analysis Center/California Institute of Technology, funded by the National Aeronautics and Space Administration and the National Science Foundation. The SIMBAD database has been consulted for the bibliography.

\bibliographystyle{elsarticle-harv}


\end{document}